\documentclass[useAMS,usenatbib]{iau}
\usepackage{graphicx}
\usepackage{hyperref}
\usepackage{aas_macros}

\title[Angular Momentum of Galaxies] 
{Angular momentum evolution of galaxies: the perspective of hydrodynamical simulations}

\author[Claudia del P. Lagos]   
{Claudia del P. Lagos$^{1,2,3}$
 }

\affiliation{$^{1}$International Centre for Radio Astronomy Research (ICRAR), M468, University of Western Australia, 35 Stirling Hwy, Crawley, WA 6009, Australia.\\ [\affilskip]
$^{2}$ARC Centre of Excellence for All Sky Astrophysics in 3 Dimensions (ASTRO 3D).\\[\affilskip]
$^{3}$Cosmic Dawn Center (DAWN), Niels Bohr Institute, University of Copenhagen, Copenhagen, Denmark\\email: {\tt claudia.lagos@icrar.org}}

\pubyear{2018}
\volume{xxx}  
\pagerange{XXX}
\jname{Galactic Angular Momentum}
\editors{XXX}
\begin{document}

\maketitle

\begin{abstract}
Until a decade ago, galaxy formation simulations were unable to simultaneously reproduce the
observed angular momentum (AM) of galaxy disks and bulges. Improvements in the
interstellar medium and stellar feedback modelling, together with advances in
computational capabilities, have allowed the current generation of cosmological
galaxy formation simulations to reproduce the diversity of AM and
morphology that is observed in local galaxies. In this review I discuss
where we currently stand in this area from the perspective of hydrodynamical
simulations, specifically how galaxies gain their AM, and the
effect galaxy mergers and gas accretion have on this process. I discuss results which suggest that a revision of the classical theory of
disk formation is needed, and finish by discussing what the current challenges are.
\end{abstract}

\firstsection 
\section{Introduction}

The formation of galaxies can be a highly non-linear process, with many physical mechanisms interacting
simultaneously. Despite all that potential complexity,
early studies of galaxy formation stressed the importance of three quantities
to describe galaxies: mass, $M$, energy, $E$, and angular momentum (AM), $J$ (e.g. Peebles 1969)\nocite{Peebles69}; one can alternatively define the specific AM, $j\equiv J/M$, which
contains information on the scale length and rotational velocity of a system. It is therefore intuitive to expect the relation between $j$ and $M$ to contain fundamental information.

%

\begin{figure}
\begin{center}
 \includegraphics[width=0.7\textwidth]{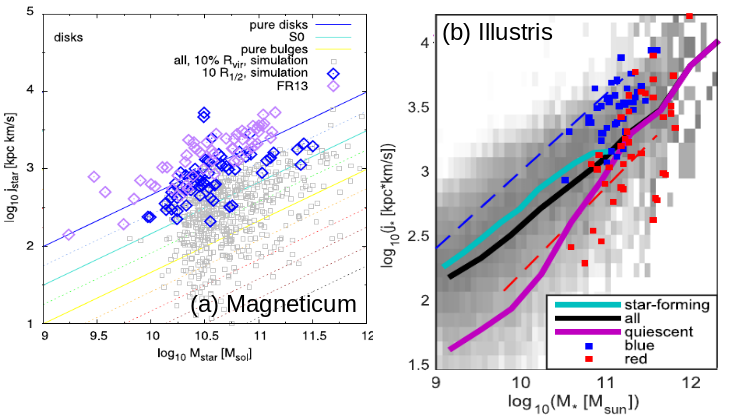}
  \includegraphics[width=0.83\textwidth]{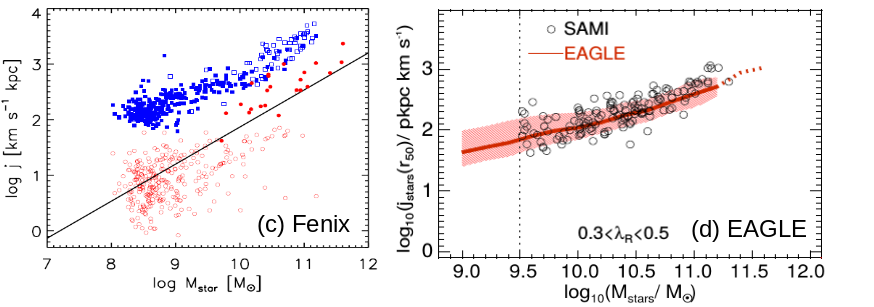}
\vspace*{-0.1 cm}\caption{The $z=0$ $j_{\rm stars}-M_{\rm stars}$ relation for several simulations: Magneticum (Teklu et al. 2015), Illustris (Genel et al. 2015), Fenix (Pedrosa \& Tissera 2015) and EAGLE (Lagos et al. 2017). Panels (a), (b) and (c) show the total $j_{\rm stars}$ compared to the observational measurements of Fall  \& Romanowsky (2013), while panel (d) shows $j_{\rm stars}$ measured within an effective radius and compares with SAMI observations (Cortese et al. 2016).}
\label{jmsims}
\end{center}
\end{figure}
In recent years, Integral field spectroscopy (IFS) is opening a new window to explore galaxy kinematics and its connection to galaxy formation and evolution, with IFS based measurements of the $j_{\rm stars}-M_{\rm stars}$ relations being reported in the local (e.g. Cortese et al. 2016\nocite{Cortese16}) and high-z Universe (Burkert et al. 2016\nocite{Burkert16}; Swinbank et al. 2017\nocite{Swinbank17}; Harrison et al. 2017\nocite{Harrison17}). The last decade has also been a golden one for cosmological hydrodynamical simulations, with the first large cosmological volumes, with high enough resolution to study the internal structure of galaxies being possible (Schaye et al. 2015; Dubois et al. 2016; Vogelsberger et al. 2014; Pillepich et al. 2018)\nocite{Schaye14,Dubois16,Vogelsberger14,Pillepich17}. These simulations have been able to overcome the catastrophic loss of AM, which refers to the problem of 
galaxies being too low $j$ compared to observations (Steinmetz \& Navarro 1999; Navarro \& Steinmetz 2000)\nocite{Steinmetz99,Navarro00} and over-cooling problem.
This problem was solved by improving the spatial resolution, adopting j conservation numerical schemes, and including efficient feedback
(e.g.Kaufmann et al. 2007\nocite{Kaufmann07}; Zavala et al. 2008\nocite{Zavala08}; Governato et al. 2010\nocite{Governato10}; Guedes et al. 2011\nocite{Guedes11}\nocite{Emsellem07}; DeFelippis et al. 2017\nocite{DeFelippis17}).
Fig.~\ref{jmsims} shows examples of several cosmological simulations {which have been} shown to reproduce the observed $j_{\rm stars}-M_{\rm stars}$ relation for $z=0$ galaxies\nocite{Teklu15,Genel15,Pedrosa15,Lagos16b,Lagos17}. In addition to those above, there are several cosmological zoom-in simulations that have shown the same level of success (e.g. Wang et al. 2018\nocite{Wang18}; El-Badry et al. 2018\nocite{El-Badry18}). 

The level of agreement of Fig.~\ref{jmsims} gives us {assurance} that we can use these simulations to study how the AM of galaxies is gained/lost throughout the process of galaxy formation and evolution. The left panel of Fig.~\ref{jevoL17} shows an example of the morphologies of simulated galaxies in the $j_{\rm stars}-M_{\rm stars}$ plane at $z=0$ in the EAGLE simulations. There is a clear correlation between a galaxy's morphology and its kinematics, as seen in observations (e.g. Cortese et al. 2016, Fall \& Romanowski 2018). An important caveat, however, is that most, if not all, the simulations of Fig.~\ref{jmsims} (and those with similar specifications) are currently unable to form very thin disks (with ellipticities $\gtrsim 0.7-0.8$) due to the insufficient resolution and simplistic interstellar medium (ISM) models. The latter prevent us from obtaining a realistic vertical structure of disks (see discussion in Lagos et al. 2018a\nocite{Lagos18b}). 

\vspace{-0.5cm}\section{How galaxies gain their angular momentum?}

\begin{figure}
\begin{center}
 \includegraphics[width=0.87\textwidth]{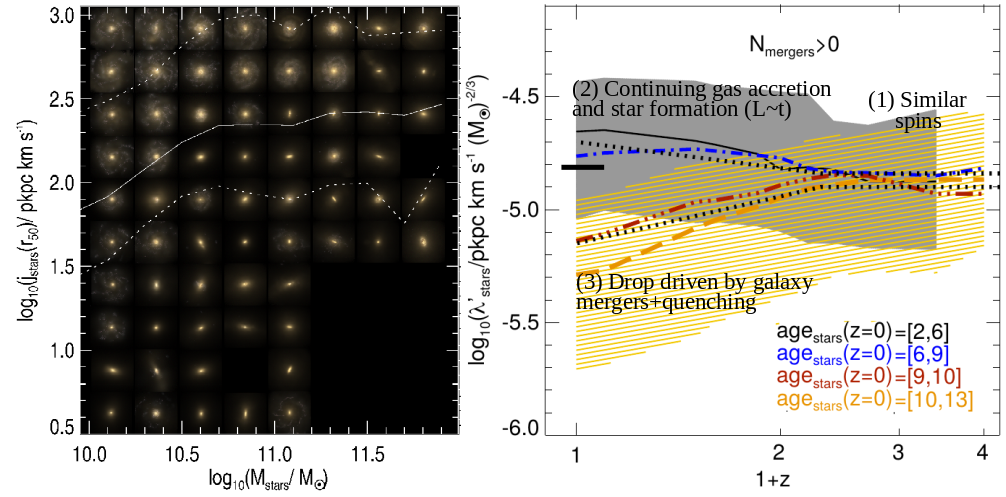}
\vspace*{-0.2 cm}\caption{{\it Left panel:} From Lagos et al. (2018b). Synthetic gri optical images of randomly selected galaxies in the $\rm j_{\rm stars}(r_{50})-M_{\rm stars}$ plane at $z=0$. The solid and dashed lines show the median and the $16^{\rm th}-84^{\rm th}$ percentile range. {\it Right panel:} Adapted from Lagos et al. (2017). $\lambda^{\prime}_{\rm stars}$ as a function of redshift for galaxies that by $z=0$ have different stellar ages and that have had at least one merger with a mass ratio $\ge 0.1$. The shaded regions show the $16^{\rm th}-84^{\rm th}$ percentile ranges for the lowest and highest age bins. At $z\gtrsim 1.2$, galaxies have similar spins, while diverging dramatically at lower redshifts. Lagos et al. (2017) identify three critical features described in the figure.}
\label{jevoL17}
\end{center}
\end{figure}

\begin{figure}
\begin{center}
 \includegraphics[width=0.87\textwidth]{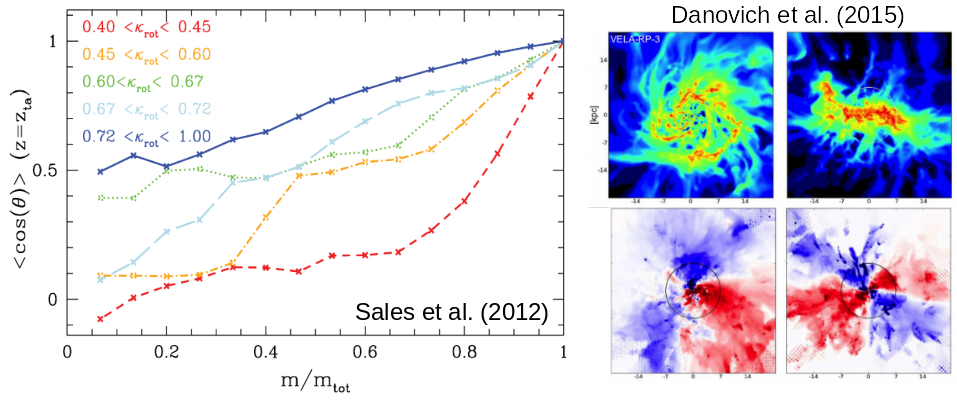}
\vspace*{-0.2 cm}\caption{{\it Left panel:} From Sales et al. (2012). The angle between the AM vector enclosed within a given mass fraction ($x$-axis) and the total spin of the system, measured at the time of maximum expansion of the halo, for galaxies that by $z=0$ have different fractions of kinetic energy invested in rotation ($\kappa_{\rm rot}$; the higher the $\kappa_{\rm rot}$ the more rotation dominates). This figure shows that alignments between the halo and galaxy are key to facilitate the formation of disks. {\it Right panel:} From Danovich et al. (2015). Face-on and edge-on density (top) and normalised torque (bottom) maps of a simulated $z\approx 2$ galaxy. The figure shows that the quadrupole torque pattern expected in the idealized case of the formation of a thin disk is seen in both orientations, meaning that the galaxy is being torqued in all three axes.}
\label{Alignments}
\end{center}
\end{figure}


Lagos et al.~(2017) used the EAGLE simulations (Schaye et al.~2015) to study how the stellar spin of galaxies at $z=0$ evolved depending on their stellar population age and merger history. The stellar spin in this case was defined as a pseudo spin $\lambda^{\prime}_{\rm stars}=j_{\rm stars}(r_{\rm 50}) / M^{2/3}_{\rm stars}$, with $j_{\rm stars}(r_{\rm 50})$ being the stellar specific AM within one effective radius. The power-law index $2/3$ comes from the predicted j-mass relation of dark matter halos (Fall~1983). The right panel of Fig.~\ref{jevoL17} shows the evolution of  $\lambda^{\prime}_{\rm stars}$ of $z=0$ galaxies with different stellar ages that have had at least $1$ galaxy merger. Progenitors display indistinguishable kinematics at $z\gtrsim 1$ despite their $z=0$ descendants being radically different ((1) in the right panel of Fig.~\ref{jevoL17}). Similarly, Penoyre et al. (2017)\nocite{Penoyre17} found using Illustris that the progenitors of $z=0$ early-type galaxies that are slow and fast rotators, had very similar properties at $z\gtrsim 1$ (see also Choi et al. 2017\nocite{Choi17} for an example using the Horizon-AGN simulation). The evolution of $\lambda^{\prime}_{\rm stars}$ diverges dramatically at $z\lesssim 1$, in which galaxies that by $z=0$ have young stellar populations, {grow} their disks efficiently due to the continuing gas accretion and star formation ((2) in Fig.~\ref{jevoL17}); galaxies that by $z=0$ have old stellar populations went through active spinning down, due to the effects of dry galaxy mergers and quenching ((3) in Fig.~\ref{jevoL17}; discussed in more detail in $\S$~\ref{mergerssec}). 

The reason why continuing gas accretion drives spinning up is because the AM of the material falling into halos is expected to increase linearly with time (as predicted by tidal torque theory; Catelan \& Theuns 1996\nocite{Catelan96a}). El-Badry et al. (2018) explicitly demonstrated this using the FIRE simulations. Garrison-Kimmel et al. (2018\nocite{Garrison-Kimmel18}) also using FIRE, in fact argued that the most important predictor of whether a disk will be formed by $z=0$ is the halo gas $j$ by the time the galaxy has formed half of its stars. These simulations thus suggest that {\it the later the accretion the more efficient the spinning up}. 

Simulations suggest the critical transition at $z\approx 1$ { is } driven by a change in the main mode of gas accretion onto galaxies, from filamentary accretion to gas cooling from a hydrostatic halo (e.g. Garrison-Kimmel et al. 2018). The latter seems to be key in facilitating alignments between the accreting gas and the galaxy, while the former is by nature more stochastic. Sales et al. (2012)\nocite{Sales12} showed that galaxies that by $z=0$ are more rotation-dominated formed in halos that had the inner/outer parts better aligned (see left panel in Fig.~\ref{Alignments}). 
Similarly, Stevens et al. (2017)\nocite{Stevens16b} showed that significant AM losses of the cooling gas to the hot halo are seen in cases where the hot halo is more misaligned with the galaxy. On the other hand, filamentary accretion at $z\gtrsim 1$ is not as efficient in spinning {up} galaxies mostly because gas filaments arrive from different directions (e.g. Walker et al. 2017\nocite{Welker17}), typically causing torques to act in all three axes of a galaxy (Danovich et al. 2015\nocite{Danovich15}; see left panel in Fig.~\ref{Alignments}). The latter is intimately connected to high-z disks being turbulent and highly disturbed. 

An important result that is robust to the details of the simulation being used, is that the circum-galactic medium (CGM) seems to have a specific AM in excess to that of the halo by factors of $3-5$ (Stewart et al. 2017\nocite{Stewart17}; Stevens et al. 2017). Stevens et al. (2017) also showed that about $50-90$\% of that excess $j$ can be lost to the hot halo in the process of gas cooling and accretion onto the galaxy. 

\vspace{-0.4cm}\section{How galaxies loose their angular momentum?}\label{mergerssec}

\begin{figure}
\begin{center}
 \includegraphics[width=0.99\textwidth]{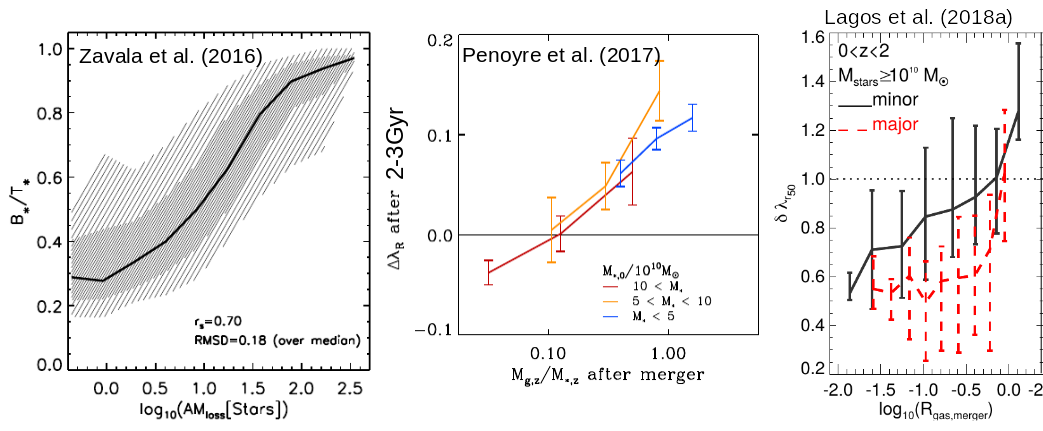}
\vspace*{-0.1 cm}\caption{{\it Left panel:} From Zavala et al. (2016). There is a strong correlation between the bulge-to-total ratio of $z=0$ galaxies ($x$-axis) and the net AM loss suffered by the galaxy stars (i.e. the difference between the maximum and $z=0$ $j_{\rm stars}$; $y$-axis). {\it Middle panel:} From Penoyre et al. (2017). Change in stellar spin parameter in galaxies (as defined in Emsellem et al. 2007) approximately $2-3$~Gyr after a merger, as a function of the gas fraction of the secondary galaxy which merged. Galaxies can become more rotation- ($\Delta(\lambda_{\rm R})>0$) or dispersion-dominated ($\Delta(\lambda_{\rm R})<0$), depending of the gas brought up by the secondary galaxy. {\it Right panel:} From Lagos et al. (2018a). Fractional change in $\lambda_{\rm R}$ as a function of the total gas-to-stellar mass ratio of the merging system, for minor (mass ratios between $0.1-0.3$) and major (mass ratios $\ge 0.3$) mergers. Gas-poor mergers are required to significantly spin galaxies down.}
\label{Mergers}
\end{center}
\end{figure}
The latest generation of hydrodynamical simulations has been able to approximately reproduce the morphological diversity of galaxies (e.g. Vogelsberger et al. 2014\nocite{Vogelsberger14}; Dubois et al. 2016\nocite{Dubois16}; see left panel in Fig.~\ref{jevoL17}).
A long-standing question is therefore how do galaxies become elliptical with $j$ significantly below disk galaxies of the same mass?

Zavala et al. (2016) used the EAGLE simulations to study the AM of galaxies and found a very strong correlation between the kinematic stellar bulge-to-total ratio and the net loss in AM of the stars that end up in galaxies at $z=0$ (left panel Fig.~\ref{Mergers}), suggesting galaxy mergers to be good candidates for the physical process behind this correlation. More recently, several authors have shown (e.g. Penoyre et al. 2017; Lagos et al. 2018a,b) that the gas fraction of the merger is one of the key parameters indicating whether the merger will lead to the primary galaxy spinning up or down. Middle and right panels of Fig.~\ref{Mergers} show the clear effect of galaxy mergers on the stellar spin parameter of galaxies in Illustris and EAGLE, respectively. 

Lagos et al. (2018b) showed that other merger parameters can have a significant effect on the $j_{\rm stars}$ structure, with high orbital $j$ and/or co-rotating mergers driving {more efficient spinning up}. However, gas fraction is the single strongest parameter that determines whether a galaxy spins up or down as a result of the merger, with the mass ratio modulating the effect. Active Galactic Nuclei feedback is key to prevent further gas accretion and the regrowth of galaxy disks in elliptical galaxies (e.g. Dubois et al. 2016). 
Early works on dry mergers (e.g. Navarro \& White 1994\nocite{Navarro94}) show that dynamical friction redistributes $j_{\rm stars}$ in a way such that most of it ends up at very large radii. {However, when} integrating over a large enough baseline, one should find $j_{\rm stars}$ converging to $j_{\rm halo}$. Using EAGLE, Lagos et al. (2018b) confirmed that to be the case: gas-poor mergers do not significantly change the {\it total} $j$ of the system, but significantly re-arrange it so that the inner parts of galaxies ($r/r_{\rm 50}<5$) become highly deficient in $j$ compared to galaxies of the same stellar mass that went through gas-rich mergers or not mergers at all. This was also seen by Teklu et al. (2015) as very deficient $j_{\rm stars}$ profiles in early-types compared to late-types at $r/R_{\rm vir}<0.2$. An important prediction of that process is that the  cumulative $j_{\rm stars}$ radial profiles of high $j$ galaxies are much more self-similar than those of galaxies with low $j$. In other words: there are few ways in which a galaxy by $z=0$ can end up with high $j$, but many pathways that lead to low $j$ (e.g. Naab et al. 2014\nocite{Naab14}, Garrison-Kimmel et al. 2018).

\vspace{-0.5cm}\section{Discussion and future prospects}

The picture that has emerged from simulations in how galaxies gain their AM is significantly more complex than the classical picture of galaxy disks forming inside out observing $j$ conservation (e.g. Mo, Mao \& White 1998\nocite{Mo98}). This complexity, however, is driven by processes that act in different directions and that tend to compensate quite efficiently, so that galaxies follow the classical disk formation model to within $50$\% (Zavala et al. 2016; Lagos et al. 2017, 2018b). 

This inevitably opens the {following} question: to what extent are we forcing $j_{\rm stars}\sim j_{\rm halo}$ in disk galaxies through the process of tuning free-parameters 
in simulations? State-of-the-art simulations tend to carefully tune their parameters to reproduce some broad statistics of galaxies, such as the stellar-halo mass relation, stellar mass function, and in some cases the size-mass relation (e.g. Crain et al. 2015\nocite{Crain15}). A consequence of such tuning may well result in this 
conspiracy: the CGM's $j$ being largely in excess of the halo's $j$, but then losing significant amounts of it while falling onto the galaxy, so that by $z=0$ disk galaxies 
have $j_{\rm stars}\sim j_{\rm halo}$. 
A possible solution for this conundrum is to perform detailed, 
high-resolution simulations of individual disk galaxies in a cosmological context and test widely different feedback mechanisms 
with the aim of understanding which conditions lead to $j_{\rm stars}\sim j_{\rm halo}$ and how independent the tuning of parameters 
is of the evolution of specific AM.

Another important area of research will be {the improvement of the description of} the vertical structure of disks, as current large cosmological hydrodynamical simulations struggle to form thin disks $\epsilon\gtrsim 0.7-0.8$. 
This is most likely due to the ISM and cooling modelling and resolution in these simulations being insufficient. Currently, simulations tend to force the gas to not cool down below $\approx 10^4$~K, which corresponds
to a Jeans length of $\approx 1$~kpc, much larger than the scaleheights of
disks in the local Universe.
This issue could be solved by including the formation
of the cold ISM, which necessarily means improving the resolution of the simulations significantly.

\bibliographystyle{mn2e_trunc8}
\vspace{-0.5cm}\bibliography{KinematicQuenching}

\end{document}